\documentclass{aa}
\usepackage{graphics}
\usepackage{times}
\begin{document}
\thesaurus{11(11.04.1; 11.19.2; 12.04.3)}
\title{Kinematics of the Local Universe.}
\subtitle{
X. $H_0$ from the inverse B-band Tully-Fisher relation using
diameter and magnitude limited samples
}
\author{Ekholm,~T.\inst{1,4}
\and Teerikorpi,~P.\inst{1} 
\and Theureau,~G.\inst{2,3}
\and Hanski,~M.\inst{1}
\and Paturel,~G.\inst{4}
\and Bottinelli,~L.\inst{3}
\and Gouguenheim,~L.\inst{3}}
\offprints{T.~Ekholm}
\institute{
 Tuorla Observatory, FIN-21500 Piikki\"o, Finland
\and Osservatorio Astronomico di Capodimonte, I-80 131 Napoli, Italy 
\and Observatoire de Paris-Meudon, CNRS URA1757,
 F92195 Meudon Principal CEDEX, France
\and Observatoire de Lyon, F69561 Saint-Genis Laval CEDEX, France
}
\date{received  , accepted  }
\maketitle
%
%
%
%
\begin{abstract}
We derive the value of $H_0$ using the inverse diameter and
magnitude B-band Tully-Fisher 
relations and the large all-sky sample
KLUN (5171 spiral galaxies). Our kinematical model was
that of Peebles centered at Virgo.
Our calibrator sample consisted of 15 field galaxies 
with cepheid distance moduli measured mostly with HST. 
A straightforward application of the inverse relation
yielded 
$H_0\approx 80\mathrm{\ km\,s^{-1}\,Mpc^{-1}}$  
for the diameter relation and
$H_0\approx 70\mathrm{\ km\,s^{-1}\,Mpc^{-1}}$
for the magnitude relation. 
$H_0$ from diameters is about 50 percent and
from magnitudes about 30 percent larger than 
the corresponding direct estimates
(cf. Theureau et al. \cite{Theureau97b}).
This discrepancy could not be resolved in terms
of a selection effect in $\log V_\mathrm{max}$
nor by the dependence of the zero-point on the
Hubble type.

We showed that a new, calibrator
selection bias (Teerikorpi \cite{Teerikorpi99}), is present. 
By using samples of signicificant size (N=2142
for diameters and N=1713 for magnitudes) we found 
for a homogeneous distribution of galaxies 
($\alpha=0$):
\begin{itemize}
\item $H_0=52^{+5}_{-4}\mathrm{\ km\,s^{-1}\,Mpc^{-1}}$ 
for the inverse diameter
B-band Tully-Fisher relation, and
\item $H_0=53^{+6}_{-5}\mathrm{\ km\,s^{-1}\,Mpc^{-1}}$ 
for the inverse magnitude
B-band Tully-Fisher relation.
\end{itemize}
Also $H_0$'s from a fractal distribution of galaxies 
(decreasing radial 
number density gradient $\alpha=0.8$) agree
with the direct predictions.
This is the first time when the
{\it inverse} Tully-Fisher relation clearly lends credence
to small values of the Hubble constant $H_0$
and to long cosmological distance scale consistently supported by
Sandage and his collaborators. 
\end{abstract}
%
%
%
%
\section{Introduction}
The determination of the value of the Hubble constant, $H_0$, 
is one of the classical tasks of {\it observational}
cosmology. In the framework of the expanding space paradigm it
provides a measure of the distance scale in FRW universes and
its reciprocal gives the time scale.
This problem has been approached in various ways.
A review on the recent determinations of the value of $H_0$
shows that most methods provide values at $H_0\sim55\dots75$
(for brevity we omit the units; all $H_0$ values are in
$\mathrm{\ km\,s^{-1}\,Mpc^{-1}}$): Virgo cluster yields
$55\pm7$ and clusters from Hubble diagram with relative
distances to Virgo $57\pm7$ (Federspiel et al.
\cite{Federspiel98}), type Ia supernovae give
$60\pm10$ (Branch \cite{Branch98}) or $65\pm7$ (Riess et
al. \cite{Riess98}), Tully-Fisher relation in I-band yields
$69\pm5$ (Giovanelli et al. \cite{Giovanelli97}) and
$55\pm7$ in B-band (Theureau et al. \cite{Theureau97b}, value
and errors combined from the diameter and magnitude 
relations), red giant branch tip gives $60\pm11$
(Salaris \& Cassisi \cite{Salaris98}), gravitational lens
time delays $64\pm13$ (Kundi\'c et al. \cite{Kundic97})
and the `sosies' galaxy method $60\pm10$ (Paturel
et al. \cite{Paturel98}). Sunyaev-Zeldovich effect
has given lower values, $49\pm29$ by Cooray
(\cite{Cooray98}), $47^{+23}_{-15}$ by Hughes \&
Birkinshaw (\cite{Hughes98}), but the uncertainties in
these results are large due to various systematical
effects (Cen, \cite{Cen98}). Surface brightness 
fluctuation studies provide a higher value of 
$87\pm11$ (Jensen et al., \cite{Jensen99}), but
most methods seem to fit in the range 55 - 75
stated above. An important comparison to these local
values
may be found after the cosmic microwave background
anisotropy probes (MAP and Planck) and galaxy
redshift surveys (2dF and SDSS) offer us a
multitude of high resolution data (Eisenstein
et al., \cite{Eisenstein98}). Note that most of
the errors cited here as well as given in the
present paper are $1\sigma$ errors.

The present line of research has its roots in the work of
Bottinelli et al. (\cite{Bottinelli86}), where $H_0$ was
determined using spiral galaxies in the {\it field}. They
used the direct Tully-Fisher relation (Tully \&
Fisher \cite{Tully77}): 
%
\begin{equation}
\label{E1} 
\\        M\propto\log V_\mathrm{max},
\end{equation}
where $M$ is the absolute magnitude in a given band
and $\log V_\mathrm{max}$ is the maximum rotational
velocity measured from the hydrogen 21 cm line width of each galaxy.
Gouguenheim (\cite{Gouguenheim69})
was the first to suggest that such a relation might exist
as a distance indicator.

Bottinelli et al. (\cite{Bottinelli86}) paid particular attention
to the elimination of the so-called {\it Malmquist bias}. In
general terms, the determination of $H_0$ is subject to the 
Malmquist bias of the $2^\mathrm{nd}$ kind: 
the inferred value of $H_0$ depends on the distribution 
of the derived distances $r$ for each true distance $r'$ 
(Teerikorpi \cite{Teerikorpi97}). 
Consider the expectation value of the derived distance $r$
at a given true distance $r'$:
%
\begin{equation}
\label{E2}
\\  E(r\vert r')=\int\limits_0^{\infty}\!\mathrm{d}r\,r\,P(r\vert r').
\end{equation}
The integral is done over {\it derived} distances $r$.
For example, consider a strict magnitude
limit: for each true distance the derived distances are exposed to an
upper cut-off. Hence the expectation value for the derived 
distance $r$ at $r'$ is too small and thus
$H_0$ will be overestimated.

Observationally, the direct Tully-Fisher relation takes the form:
%
\begin{equation}
\label{E3}
\\ X = \mathrm{slope}\times p + \mathrm{cst},
\end{equation}
where we have adopted a shorthand
$p$ for $\log V_\mathrm{max}$ and $X$ denotes either the
absolute magnitude $M$ or $\log D$, where $D$ labels the
absolute linear size of a galaxy in kpc. 
In the {\it direct} approach the slope is determined from
the linear regression of $X$ against $p$. The resulting
direct Tully-Fisher relation can be expressed as
%
\begin{equation}
\label{E4}
\\ E(X\vert p) = ap+b.
\end{equation}
Consider now the {\it observed} average of $X$ at each $p$, 
$\langle X\rangle_p$, as a function of the true distance. 
The limit in $x$ (the observational counterpart of $X$) 
cuts off progressively more and more of the distribution 
function of $X$ for a constant $p$. 
Assuming $X=\log D$ one finds:
%
\begin{equation}
\label{E5}
\\ \langle X\rangle_p \ge E(X\vert p),
\end{equation}
The inequality gives a practical measure of the Malmquist
bias depending primarily on $p$, $r'$, $\sigma_X$ 
and $x_\mathrm{lim}$. 
The equality holds only when the $x$-limit cuts the
luminosity function $\Phi(X)$ insignificantly.

That the direct relation is {\it inevitably} biased by its
nature forces one either to look for an unbiased subsample
or to find an appropriate correction for the bias. The
former was the strategy chosen by Bottinelli et al.
(\cite{Bottinelli86}) where the method of normalized distances
was introduced. This is the method chosen also by the KLUN project. 
KLUN ({\sl Kinematics of the Loal Universe}) is based on a
large sample, which consists of
5171 galaxies of Hubble types T=1-8 distributed on the whole 
celestial sphere (cf. e.g. Paturel \cite{Paturel94}, 
Theureau et al. \cite{Theureau97b}). 

Sandage (\cite{Sandage94a}, \cite{Sandage94b}) 
has also studied the latter approach. By recognizing that the
Malmquist bias depends not only on the imposed $x$-limit
but also on the rotational velocities and distances, he
introduced the triple-entry correction method,
which has consistently predicted values of $H_0$ supporting 
the long cosmological distance scale. As a practical
example of this approach to the Malmquist bias cf.
e.g. Federspiel et al. (\cite{Federspiel94}).

Bottinelli et al. (\cite{Bottinelli86}) found 
$H_0=72\mathrm{\, km\, s^{-1}\, Mpc^{-1}}$
using the method of normalized distances, i.e. using a sample 
cleaned of galaxies suffering from the Malmquist bias. 
This value was based on the de Vaucouleurs calibrator distances.
If, instead, the Sandage-Tammann calibrator distances were used
Bottinelli et al. (\cite{Bottinelli86})  found 
$H_0=63\mathrm{\, km\, s^{-1}\, Mpc^{-1}}$
(or  $H_0=56\mathrm{\, km\, s^{-1}\, Mpc^{-1}}$
if using the old ST calibration).
One appreciates the debilitating effect of the Malmquist
bias by noting that when it is ignored
the de Vaucouleurs calibration yields much larger values:
$H_0\sim100\mathrm{\, km\, s^{-1}\, Mpc^{-1}}$.

Theureau et al. (\cite{Theureau97b})
by following the guidelines set out by Bottinelli et
al. (\cite{Bottinelli86}) determined the value of $H_0$
using the KLUN sample.
$H_0$ was determined not only using magnitudes
but also diameters because the
KLUN sample is constructed to be complete in angular
diameters rather than magnitudes (completeness limit
is estimated to be $D_{25}=1\farcm6$).
Left with 400 unbiased galaxies (about ten times more than
Bottinelli et al. (\cite{Bottinelli86}) were able to use) reaching
up to $2000{-}3000\mathrm{\, km\, s^{-1}}$
they found using the most recent calibration based on HST
observations of extragalactic cepheids
\begin{itemize}
\item $H_0=53.4\pm 5.0\mathrm{\, km\, s^{-1}\, Mpc^{-1}}$
from the magnitude relation, and
\item $H_0=56.7\pm 4.9\mathrm{\, km\, s^{-1}\, Mpc^{-1}}$
from the diameter relation.
\end{itemize}
They also discussed in their Sect. 4.2 how these results
change if the older calibrations were used. For example,
the de Vaucouleurs calibration would increase these values
by 11 \%. We expect that a similar effect would be 
observed also in the present case.

In the present paper we ask whether the results of
Theureau et al. (\cite{Theureau97b}) could be confirmed 
by implementing the {\it inverse} Tully-Fisher relation:
%
\begin{equation}
\label{E6}
\\     p=a'X+b',
\end{equation}
This problem has special importance because of the ``unbiased''
nature that has often been ascribed to the inverse 
Tully-Fisher relation as a distance
indicator and because of the large number of galaxies available contrary to
the direct approach where one is constrained to the 
so called unbiased plateau (cf. Bottinelli et al. \cite{Bottinelli86}; 
Theureau et al. \cite{Theureau97b}). The fact that the inverse relation
has it own particular biases has received increasing attention
during the years (Fouqu\'e et al.
\cite{Fouque90}, Teerikorpi \cite{Teerikorpi90}, Willick
\cite{Willick91}, Teerikorpi \cite{Teerikorpi93},
Ekholm \& Teerikorpi \cite{Ekholm94}, Freudling et al.
\cite{Freudling95}, Ekholm \& Teerikorpi \cite{Ekholm97},
Teerikorpi et al. \cite{Teerikorpi99} and, of course, the
present paper).
%
%
%
%
\section{Outlining the approach}
As noted in the introduction the KLUN project approaches the
problem of the determination of the value of $H_0$ using
field galaxies with photometric distances. Such an approach
reduces to three steps
\begin{enumerate}
\item construction of a relative kinematical distance scale,
\item construction of a relative redshift-independent
distance scale, and
\item establishment of an absolute calibration.
\end{enumerate}
Below we comment on the first two steps. In particular we
further develop the concept of a relevant inverse slope
which may differ from the theoretical slope, but is still
the slope to be used. The third step is addressed in Sect.~6. 
It is hoped that this review clarifies the methodological basis 
of the KLUN project and also makes the notation used more familiar.
\subsection{The kinematical distance scale}
The first step takes its simplest form by assuming
the strictly linear Hubble law:
%
\begin{equation}
\label{E7}
\\ R_\mathrm{kin} = V_\mathrm{o}/H_0'
\end{equation}
where $V_\mathrm{o}$ is the radial velocity inferred from
the observed redshifts and $H_0'$ is some input value
for the Hubble constant. Because $V_\mathrm{o}$ reflects
the true kinematical distance $R_\mathrm{kin}^*$ via the
true Hubble constant $H_0^*$
%
\begin{equation}
\label{E8}
\\ R_\mathrm{kin}^* = V_\mathrm{o}/H_0^*,
\end{equation}
one recognizes that Eq.~\ref{E7} sets up a {\it relative}
distance scale:
%
\begin{equation}
\label{E9}
\\ d_\mathrm{kin} = \frac {R_\mathrm{kin}}{R_\mathrm{kin}^*}
= \frac {H_0^*}{H_0'}.
\end{equation}
In other words, $\log d_\mathrm{kin}$ is known next to 
a constant.

In a more realistic case one ought to consider also the
{\it peculiar} velocity field. In KLUN one
assumes that peculiar velocities are governed 
mainly by the Virgo supercluster. 

In KLUN the kinematical distances are inferred
from $V_\mathrm{o}$'s by implementing the spherically
symmetric model of Peebles (\cite{Peebles76}) valid
in the linear regime. In the adopted form of this
model (for the equations to be solved cf. e.g. 
Bottinelli et al. \cite{Bottinelli86}, Ekholm \cite{Ekholm96}) 
the centre of the peculiar velocity field is marked by the pair 
of giant ellipticals M86/87 positioned at some unknown true distance 
$R^*$ which is used to normalize the kinematical distance
scale: the centre is at a distance $d_\mathrm{kin}=1$.

The required cosmological velocities $V_\mathrm{cor}$
(observed velocities corrected for peculiar motions) 
are calculated as
%
\begin{equation}
\label{E10}
\\ V_\mathrm{cor} = C_1\times d_\mathrm{kin},
\end{equation}
where the constant $C_1$ defines the linear recession
velocity of the centre of the system assumed to be
at rest with respect to the quiescent Hubble flow:
%
\begin{equation}
\label{E11}
\\ C_1 = V_\mathrm{o}(\mathrm{Vir})+
V_\mathrm{inf}^\mathrm{LG}.
\end{equation}
$V_\mathrm{o}(\mathrm{Vir})$ is the presumed velocity of the
centre and $V_\mathrm{inf}^\mathrm{LG}$ is the presumed
infall velocity of the Local Group into the centre of
the system.
\subsection{The redshift-independent distances}
The direct Tully-Fisher relation is quite
sensitivite to the sampling of the luminosity function.
On the other hand, when implementing the inverse Tully-Fisher
relation (Eq.~\ref{E6}) under ideal conditions 
it does not matter how we sample $X$
(Schechter \cite{Schechter80}) in order to obtain an 
unbiased estimate for the inverse parameters and, 
furthermore, the expectation value
$E(r\vert r')$ is also unbiased (Teerikorpi \cite{Teerikorpi84}). 
{\it However, we should sample all 
$\log V_\mathrm{max}$ for each constant {\it true} $X$ in the sample}. 
This theoretical prerequisition is often tacitly assumed
in practice. For more formal treatments on the inverse relation
cf. Teerikorpi(\cite{Teerikorpi84}, \cite{Teerikorpi90}, 
\cite{Teerikorpi97}) and e.g.
Hendry \& Simmons (\cite{Hendry94}) or Rauzy \& Triay (\cite{Rauzy96}).

In the inverse approach the distance indicator is
%
\begin{equation}
\label{E12}
\\ X = A'\langle p\rangle_X+\mathrm{cst.},
\end{equation}
where $A'=1/a'$ following the notation adopted by Ekholm
\& Teerikorpi (\cite{Ekholm97}; hereafter ET97). 
The inverse regression slope $a'$ is expected to fulfill
%
\begin{equation}
\label{E13}
\\ \langle p\rangle_X \equiv E(p\vert X)=a'X+\mathrm{cst}.
\end{equation}
$\langle p\rangle_X$ is the observed average $p$ for
a given $X$. Eq.~\ref{E13} tells that in order to find
the correct $a'$ one must sample the distribution function
$\phi_X(p)$ in such a way that 
$\langle p\rangle_X=(p_0)_X$, where $(p_0)_X$ is the
central value of the underlying distribution function.
$\phi_X(p)$ is presumed to be {\it symmetric}
about $(p_0)_X$ for all $X$. 
ET97 demonstrated
how under these ideal conditions the derived $\log H_0$ 
as a function of the kinematical distance should run
horizontally as the adopted slope approaches the ideal,
theoretical slope.

In practice the parameters involved are subject to uncertainties,
in which case one should use instead of the unknown theoretical slope
a slope which we call the {\it relevant} inverse slope.
We would like to clarify in accurate terms the meaning of this slope 
which differs from the theoretical slope
and which has been more heuristically discussed by 
Teerikorpi et al. (\cite{Teerikorpi99}). 
The difference between the theoretical and the relevant slope can be
expressed in the following formal way. Define the observed parameters
as
%
\begin{equation}
\label{E14}
\\ X_\mathrm{o}=X+\epsilon_x+\epsilon_\mathrm{kin},
\end{equation}
%
%
\begin{equation}
\label{E15}
\\ p_\mathrm{o}=p+\epsilon_p,
\end{equation}
where $X$ is inferred from $x$ with a measurement error $\epsilon_x$
and the kinematical distance $d_\mathrm{kin}$ has an error $\epsilon_\mathrm{kin}$ due to
uncertainties in the kinematical distance scale.
$\epsilon_p$ is the observational error on $p$.
The theoretical slope $a'_\mathrm{t}$ is\footnote
{We make use of the
formal definition of the slope of the linear regression of $y$
against $x$ with
%
\begin{equation}
\label{E16}
\\ \mathrm{Cov}(x,y) = 
\frac{\sum(x-\langle x\rangle)(y-\langle y\rangle)}{(N-1)}.
\end{equation}
}
%
\begin{equation}
\label{E17}
\\ a'_\mathrm{t}=
\frac {\mathrm{Cov}(X,p)}{\mathrm{Cov}(X,X)},
\end{equation}
while the observed slope is
%
\begin{equation}
\label{E18}
\\ a'_\mathrm{o} =
\frac {\mathrm{Cov}(X_\mathrm{o},p_\mathrm{o})}{\mathrm{Cov}
(X_\mathrm{o},X_\mathrm{o})}
\sim
\frac {\mathrm{Cov}(X,p)+\mathrm{Cov}(\epsilon_x+\epsilon_\mathrm{kin},
\epsilon_p)}
{\mathrm{Cov}(X,X)+\sigma^2_x+\sigma^2_\mathrm{kin}}
\end{equation}
We call the slope $a'_\mathrm{o}$ relevant if it verifies for 
{\it all} $X_\mathrm{o}$
(Eq.~\ref{E13})
%
\begin{equation}
\label{E19}
\\   \langle p_\mathrm{o}\rangle_{X_\mathrm{o}}=E(p\vert X_\mathrm{o})=
a'_\mathrm{o}X_\mathrm{o}+\mathrm{cst}.
\end{equation}
This definition means that 
the average observed value of $p_\mathrm{o}$ at
each fixed value of $X_\mathrm{o}$ (derived from observations and
the kinematical distance scale) is correctly predicted by
Eq.~\ref{E19}.
Note also that in the case of diameter relation, $\epsilon_x$,
$\epsilon_\mathrm{kin}$ and $\epsilon_p$ are only weakly correlated.
Thus the difference between the relevant slope and the theoretical
slope is dominated by $\sigma_x^2+\sigma_\mathrm{kin}^2$. In the
special case where the galaxies are in one cluster (i.e. at the same
true distance), the dispersion $\sigma_\mathrm{kin}$
vanishes. In order to make the relevant slope more
tangible we demonstrate in Appendix A how 
it indeed is the one to be
used for the determination of $H_0$. 

Finally, also selection in $p$ and type effect may affect the
derived slope making it even shallower. 
Theureau et al. (\cite{Theureau97a})
showed that a type effect exists seen as
degenerate values of $p$ for each constant linear diameter $X$.
Early Hubble types rotate faster than late types. In addition,
based on an observational program of 2700 galaxies with the
Nan\c{c}ay radiotelescope, Theureau (\cite{Theureau98}) 
warned that the detection
rate in HI varies continuosly from early to late types and that on
average $\sim 10\%$ of the objects remain unsuccessfully observed.
Influence of such a selection, which concerns principally the extreme
values of the distribution function $\phi(p)$, 
was discussed analytically by 
Teerikorpi et al. (\cite{Teerikorpi99}).
%
%
%
%
%
%
\section{A straightforward derivation of $\log H_0$}
\subsection{The sample}
KLUN sample is -- according to
Theureau et al. (\cite{Theureau97b}) -- complete up to
$B_\mathrm{T}^\mathrm{c}=13\fm25$, where $B_\mathrm{T}^\mathrm{c}$
is the corrected total B-band magnitude and down to
$\log D_{25}^\mathrm{c}=1.2$, where $D_{25}^\mathrm{c}$ is
the corrected angular B-band diameter.
The KLUN sample was subjected
to exclusion of low-latitude ($\vert b\vert\ge15\degr$) 
and face-on ($\log R_{25}\ge0.07$) galaxies.
The centre of the
spherically symmetric peculiar velocity field was positioned
at $l=284\degr$ and $b=74\degr$. The constant $C_1$
needed in Eq.~\ref{E10} for cosmological velocities
was chosen to be $1200\mathrm{\, km\, s^{-1}}$ with
$V_\mathrm{o}(\mathrm{Vir})=980\mathrm{\, km\, s^{-1}}$
and $V_\mathrm{inf}^\mathrm{LG}=220\mathrm{\, km\, s^{-1}}$
(cf. Eq.~\ref{E11}). After the exclusion of triple-valued
solutions to the Peebles' model and
when the photometric completeness limits cited were imposed on
the remaining sample one was left with 1713 galaxies for
the magnitude sample and with 2822 galaxies for the diameter
sample.
\subsection{The inverse slopes and calibration of zero-points}
Theureau et al. (\cite{Theureau97a}) 
derived a common inverse diameter slope $a'\approx0.50$ 
and inverse magnitude slope $a'\approx-0.10$ for 
all Hubble types considered i.e. T=1-8.
These slopes were also shown to obey a
a simple mass-luminosity model
(cf. Theureau et al. \cite{Theureau97a}).
%
%
\begin{figure}
\resizebox{\hsize}{!}{\includegraphics[211,195][460,384]{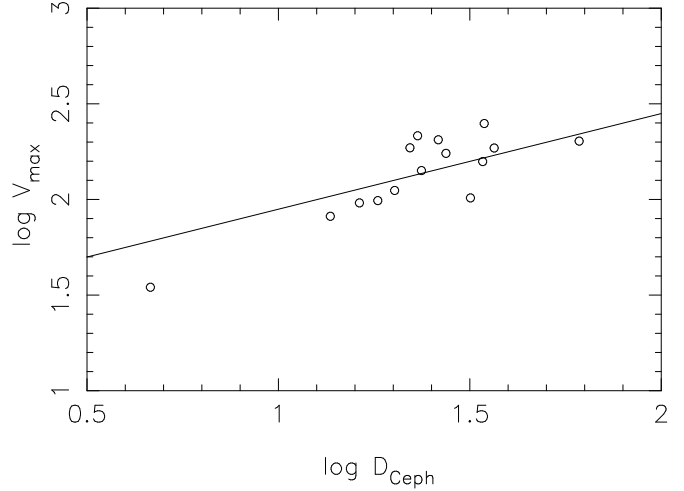}}
\caption{
The slope $a'=0.50$ forced to the calibrator 
sample with Cepheid distances yielding $b'_{\mathrm{cal}}=1.450$,
when no type corrections were made.
}
\label{F1}
\end{figure}
%
%
%
\begin{figure}
\resizebox{\hsize}{!}{\includegraphics[211,195][460,384]{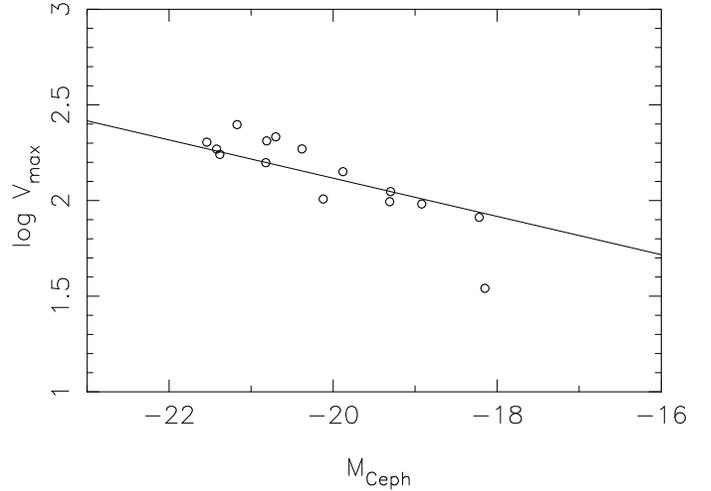}}
\caption{
The slope $a'=-0.10$ forced to the calibrator 
sample with Cepheid distances yielding $b'_\mathrm{cal}=0.117$,
when no type corrections were made.
}
\label{F2}
\end{figure}
With these estimates for the inverse
slope the relation can be calibrated. 
At this point of derivation we ignore the 
effects of type-dependence and possible selection in 
$\log V_\mathrm{max}$. 
The calibration was
done by forcing the slope to the calibrator sample 
of 15 field galaxies with cepheid distances,
mostly from the HST programs (Theureau et al.
\cite{Theureau97b}, cf. their Table~1.).
The absolute zero-point is given by
%
\begin{equation}
\label{E20}
\\   b'_\mathrm{cal}=\frac
{\sum {(\log V_\mathrm{max}-a'X)}}{N_\mathrm{cal}},
\end{equation}
where the adopted inverse slope $a'=0.50$ yields $b'_\mathrm{cal}=1.450$
and $a'=-0.10$ $b'_\mathrm{cal}=0.117$. In Fig.~\ref{F1} we show
the calibration for the diameter relation and in Fig.~\ref{F2}
for the magnitude relation.  
\subsection{$H_0$ without type corrections}
%
%
%
%
\begin{figure}
\resizebox{\hsize}{!}{\includegraphics[211,123][460,507]{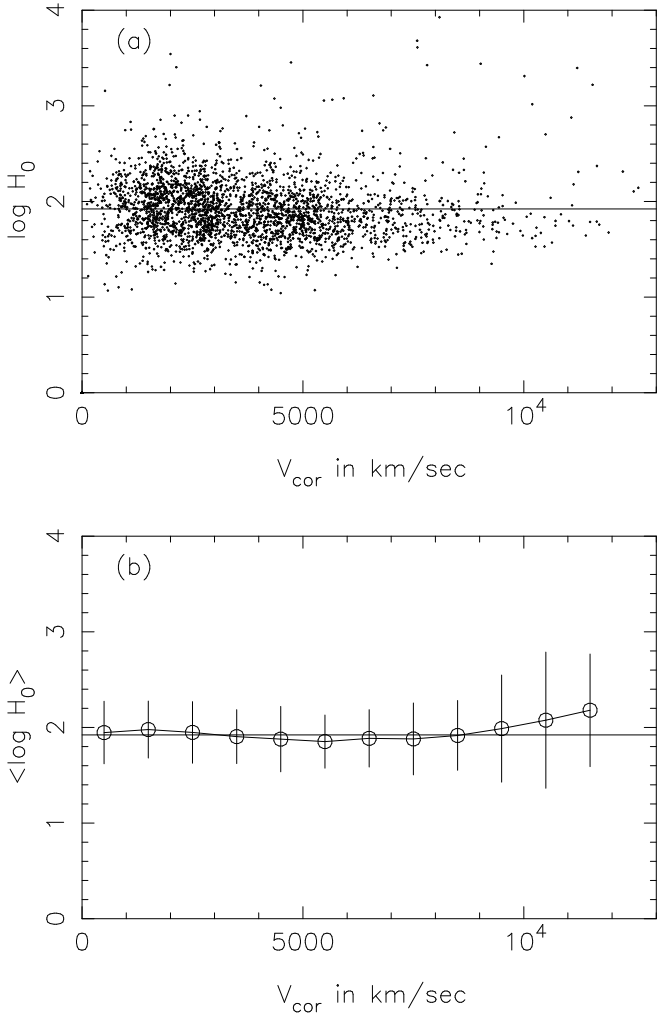}}
\caption{Panel (a): The $\log H_0$ vs. $V_\mathrm{cor}$ diagram for the
calibrated inverse Tully-Fisher relation $\log V_\mathrm{max}=
0.50\log D+1.450$. The horizontal solid line corresponds
to the average value $\langle\log H_0\rangle=1.92$. Panel (b): 
the average values $\langle\log H_0\rangle$ (circles) are shown as
well as the average of the whole sample. The averages were calculated 
for velocity bins of size $1000\mathrm{\ km\,s^{-1}}$. Total number
of points used was $N=2822$.}
\label{F3}
\end{figure}
%
%
%
\begin{figure}
\resizebox{\hsize}{!}{\includegraphics[211,123][460,507]{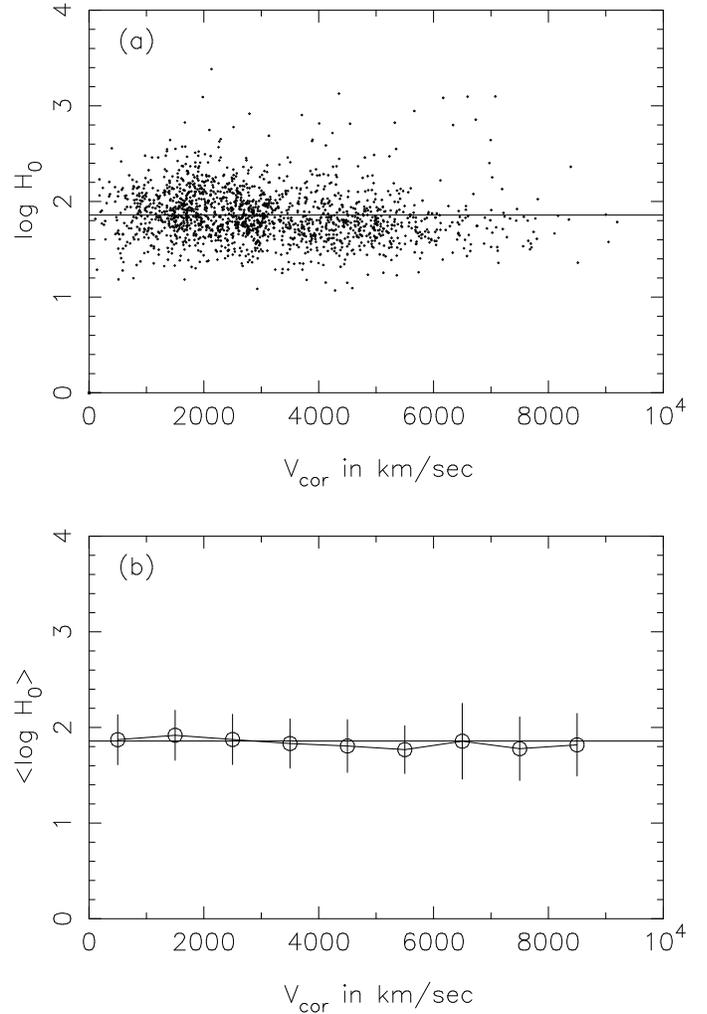}}
\caption{The sample imposed to the strict magnitude limit
$B_\mathrm{T}^\mathrm{c}=13\fm25$ (N=1713). The forced
solution yields
$\langle\log H_0\rangle=1.857$ or 
$H_0=71.9\mathrm{\, km\, s^{-1}\, Mpc^{-1}}$. 
}
\label{F4}
\end{figure}
ET97 discussed in some detail problems which hamper the determination
of the Hubble constant $H_0$ when one applies the inverse Tully-Fisher
relation. They concluded that
once the relevant inverse slope is found, the 
average $\langle\log H_0\rangle$ shows no tendencies as a function
of the distance. Or, in terms of the method of normalized distances
of Bottinelli et. al. (\cite{Bottinelli86}), 
the unbiased plateau extends to all
distances. ET97 also noted how one might simultaneously fine-tune
the inverse slope and get an unbiased estimate for $\log H_0$.
The resulting $\log H_0$ vs. kinematical distance diagrams 
for the inverse diameter relation is given in
Fig.~\ref{F3} and for the magnitude relation in Fig.~\ref{F4}. 
Application of the parameters given in the previous section yield
$\langle\log H_0\rangle=1.92$ correponding to 
$H_0=83.2\mathrm{\ km\,s^{-1}\,Mpc^{-1}}$ for the diameter sample
and
$\langle\log H_0\rangle=1.857$ or 
$H_0=71.9\mathrm{\, km\, s^{-1}\, Mpc^{-1}}$
for the magnitude sample. 
These averages are shown as horizontal, solid
straight lines. In panels (a) individual points are plotted and
in panels (b) the averages for bins of 
$1000\mathrm{\, km\, s^{-1}}$ are given as circles.

Consider first the diameter relation. One clearly
sees how the average follows a 
horizontal line up to $9000\mathrm{\ km\,s^{-1}}$.
At larger distances, the observed behaviour of
$\langle H_0\rangle$ probably
reflects some selection in $\log V_\mathrm{max}$ in the sense that 
there is an upper cut-off value for $\log V_\mathrm{max}$.
Note also the mild downward tendency between $1000\mathrm{\, km\, s^{-1}}$ 
and $5000\mathrm{\, km\, s^{-1}}$. 
Comparison of Fig.~\ref{F4} with Fig.~\ref{F3} 
shows how $\langle\log H_0\rangle$ from magnitudes
and diameters follow each other quite well as expected
(ignoring, of course, the vertical shift in the averages). 
Note how the growing tendency of 
$\langle\log H_0\rangle$ beyond $9000\mathrm{\, km\, s^{-1}}$ 
is absent in the magnitude sample
because of the limiting magnitude:
the sample is less deep. This suggests that
the possible selection bias in $\log V_\mathrm{max}$ 
does not affect the magnitude sample.

One might, by the face-value,
be content with the slopes adopted as well as with
the derived value of $H_0$. The observed behaviour is what ET97
argued to be the prerequisite for an unbiased estimate for
the Hubble constant: non-horizontal trends disappear.
It is -- however -- rather disturbing to note that the values of 
$H_0$ obtained via this straightforward application of the inverse 
relation are significantly {\it larger} than those reported
by Theureau et al. (\cite{Theureau97b}). The inverse diameter
relation predicts some
50 percent larger value and the magnitude relation some 30
percent larger value than the corresponding direct relations.
In what follows, we try to understand this discrepancy.
%
%
%
%
\section{Is there selection in $\log V_\mathrm{max}$?}
The first explanation coming to mind is that the apparently
wellbehaving slope $a'= 0.5$ ($a'=-0.1$) is incorrect because
of some selection effect and is thus {\it not} relevant in the
sense discussed in Sect.~2.2 and in Appendix A. 
The relevant slope brings about
an unbiased estimate for the Hubble parameter (or the Hubble 
constant if one possesses an ideal calibrator sample) {\it if} 
the distribution function of $\log V_\mathrm{max}$, $\phi(p)_X$,
is completely and correctly sampled for each $X$. 
Fig.~\ref{F3} showed some preliminary indications that 
this may not be the case as regards the diameter sample. 

Teerikorpi (\cite{Teerikorpi99}) discussed the effect and
significance of a strict upper and/or lower cut-off on
$\phi(p)_X$. For example, an upper cut-off in $\phi(p)_X$ 
should yield a too large value of $H_0$ and, furthermore, 
a too shallow slope. Their analytical calculations given
the gaussianity of $\phi(p)_X$ show that this kind of selection 
effect has only a minuscule affect unless the cut-offs are
considerable. Because the selection does not seem to be
significant, we do not expect much improvement in $H_0$.
%
%
\begin{figure}
\resizebox{\hsize}{!}{\includegraphics[211,195][460,384]{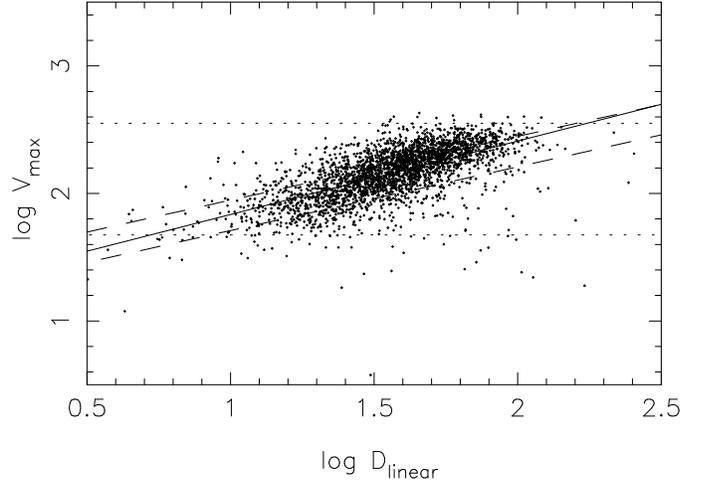}}
\caption{The inverse Tully-Fisher diagram for the sample used
in the analysis. The solid line refers to a linear regression
of $a'=0.576$ and $b'=1.256$. The dashed lines give the forced
solutions with $a'=0.50$ for Hubble types 1 with $b'=1.448$
and 8 with $b'=1.209$
The dotted lines at 
$\log V_\mathrm{max}=2.55$ and $\log V_\mathrm{max}=1.675$ are intended
to guide the eye. At least the upper cut-off is quite conspicuous.}
\label{F5}
\end{figure}

There is, however, another effect which may alter the slope. 
As mentioned in Sect.~2.2 the type-dependence of the zero-point
should be taken into account. Because the selection function
may depend on the morphological type it also affects the type
corrections. This is clearly seen when one considers how the
type corrections are actually calculated. As in Theureau et al. 
(\cite{Theureau97b}) galaxies are shifted to a common Hubble type 6 
by applying a correction term $\Delta b' = b'(T)-b'(6)$ to individual 
$\log V_\mathrm{max}$ values, where
%
\begin{equation}
\label{E21}
\\ b'(T) = 
\langle \log V_\mathrm{max}\rangle_T-a'\langle X\rangle_T. 
\end{equation}
Different morphological types do not have identical spatial occupation,
which is shown in Fig.~\ref{F5} for Hubble types 1 and 8 as dashed
lines corresponding to forced solutions using the common slope
$a'=0.5$. The strict upper and lower cut-offs would influence the
extreme types more.
Hence we must first more carefully see if the samples
suffer from selection in $\log V_\mathrm{max}$ 

The inverse Tully-Fisher diagram for the diameter sample
is given in Fig.~\ref{F5}. The least squares fit 
($a'=0.576$, $b'=1.259$) is shown as a solid line.
One finds evidence for both an
upper and lower cut-off in the $\log V_\mathrm{max}$-distribution,
the former being quite conspicuous.
The dotted lines are positioned
at $\log V_\mathrm{max}=2.55$ and $\log V_\mathrm{max}=1.675$ 
to guide the eye. 
Fig.~\ref{F5} hints
that the slope $a'=0.5$ adopted in Sect.~3 may not be impeccable
and thus questions the validity of the ``na\"{\i}ve" 
derivation of $H_0$ at least in the case of the diameter sample.
%
%
\begin{figure}
\resizebox{\hsize}{!}{\includegraphics[211,195][460,384]{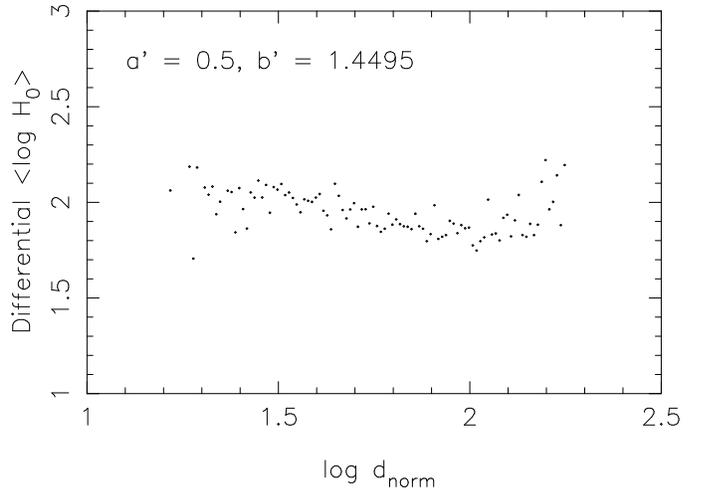}}
\caption{The differential behaviour of $\langle\log H_0\rangle$ as 
a function of the normalized distances.The inverse parameters were
$a'=0.5$ and $b'=1.450$.}
\label{F6}
\end{figure}
%
%
%
\begin{figure}
\resizebox{\hsize}{!}{\includegraphics[211,195][460,384]{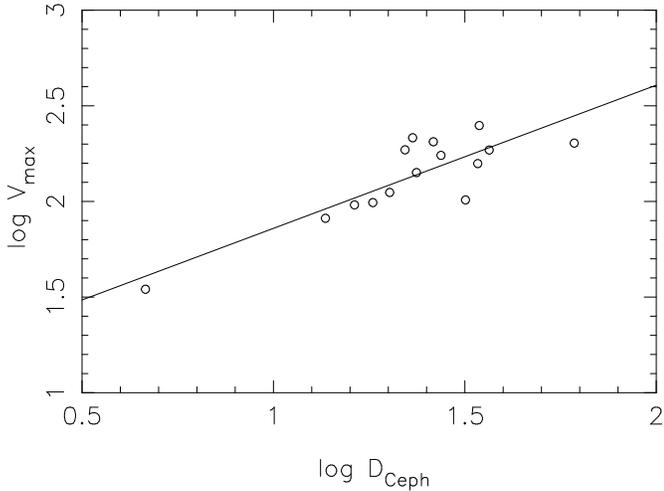}}
\caption{A straightforward linear regression applied to the calibrator
sample yielding $a'= 0.749$ and $b'= 1.101$.}
\label{F7}
\end{figure}
%
%
%
%
\begin{figure}
\resizebox{\hsize}{!}{\includegraphics[211,195][460,384]{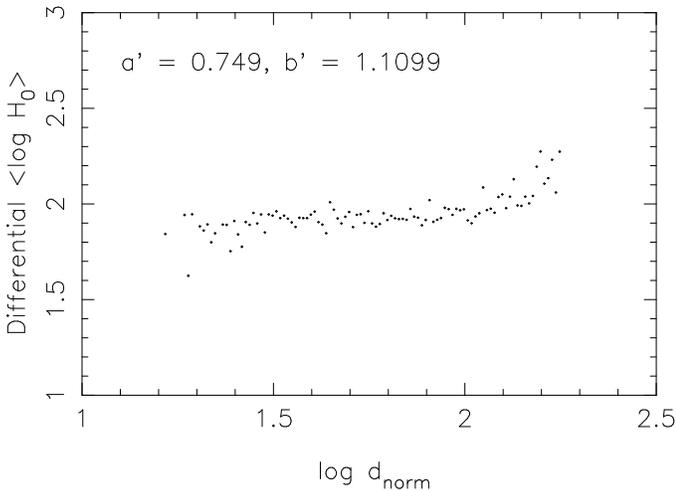}}
\caption{ As Fig.~\ref{F6}, but now the parameters
$a'= 0.749$ and $b'= 1.101$ were used.
One can see how the downward tendency between
$\log d_\mathrm{norm}\sim1.45$ and $\log d_\mathrm{norm}\sim2$ has
disappeared. Also cf. Fig. 2 in Teerikorpi et al. 
(\cite{Teerikorpi99}).}
\label{F8}
\end{figure}

In the case of diameter samples,
Teerikorpi et al. (\cite{Teerikorpi99}) 
discussed how the cut-offs should demonstrate themselves in 
a $\log H_0$ vs. $\log d_\mathrm{norm}$ diagram,
where $\log d_\mathrm{norm}=\log D_{25}+\log d_\mathrm{kin}$, 
which in fact is the log of $D_\mathrm{linear}$ next to a constant. 
We call $d_\mathrm{norm}$ ``normalized" in analogy to the
method of normalized distances, where the kinematical distances
were normalized in order to reveal the underlying bias. That is
exactly what is done also here.

Consider the differential behaviour of 
$\langle\log H_0\rangle$ as a function of the normalized distance.
Differential average $\langle\log H_0\rangle$  was calculated as
follows. The abscissa was divided into intervals of 0.01 starting
at minimum $\log d_\mathrm{norm}$ in the sample. If a bin contained
at least 5 galaxies the average was calculated.
In Fig.~\ref{F6}. the inverse parameters $a'=0.5$ and $b'=1.450$ were
used. It is seen that around $\log d_\mathrm{n}\sim 2$ the values of
$\log H_0$ have a turning point as well as at $\log d_\mathrm{n}\sim 1.45$.
The most striking feature is -- however -- the general decreasing
tendency of $\log H_0$ between these two points.
Now, according to ET97, a downward tendency of $\log H_0$
as a function of distance corresponds to $A/A'>1$, i.e. the adopted 
slope A is too shallow ($A'$ is the relevant slope).

Closer inspection of Fig.~\ref{F1} shows that a steeper slope
might provide a better fit to the calibrator sample. One is thus
tempted to ask what happens if one adopts for the field sample
the slope giving the best fit to the calibrator sample. As such
solution we adopt the straightforward linear regression yielding
$a'= 0.749$ and $b'= 1.101$ shown in Fig.~\ref{F7}. 
It is interesting to note that when these parameters are used
the downward tendency between
$\log d_\mathrm{norm}\sim1.45$ and $\log d_\mathrm{norm}\sim2$ 
disappears as can be seen in Fig.~\ref{F8}. From hereon we refer
to this interval as the ``unbiased inverse plateau". The value
of $\log H_0$ in this plateau is still rather high.
%
\begin{figure}
\resizebox{\hsize}{!}{\includegraphics[211,195][460,384]{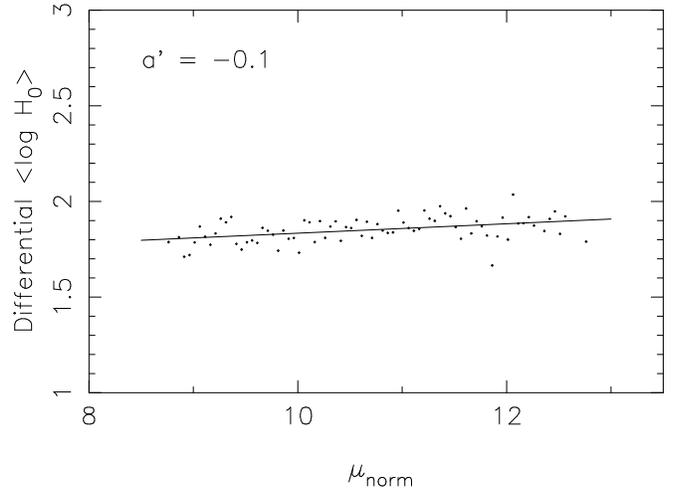}}
\caption{
The differential $\langle \log H_0\rangle$ vs. $\mu_\mathrm{norm}$
diagram. One finds no indication of a selection in
$\log V_\mathrm{max}$. The adopted slope ($a'=-0.10$) appears
to be incorrect.
}
\label{F9}
\end{figure}

In the case of the magnitude sample we study the behaviour
of the differential average $\langle \log H_0\rangle$ as
a function of a ``normalized" distance modulus:
%
\begin{equation}
\label{E22}
\\ \mu_\mathrm{norm} = B_\mathrm{T}^\mathrm{c}-5\log d_\mathrm{kin}.
\end{equation}
The $\mu_\mathrm{norm}$ axis was divided into intervals of 0.05
and again, if in a bin is more than five points the average is 
calculated. As suspected in the view of Fig.~\ref{E4}., 
Fig.~\ref{E9} reveals no significant indications of a selection in
$\log V_\mathrm{max}$. The points follow quite well the straight
line also shown. The line however is tilted telling us that
the input slope $a'=-0.10$ may not be the relevant one.

As already noted the type corrections may have some influence
on the slopes.
In the next section we derive the appropriate type corrections 
for the zero-points using galaxies residing in the unbiased
plateau ($\log d_\mathrm{norm} \in [1.45,2.0]$) for the diameter
sample and for the whole magnitude sample and rederive the
slopes.
%
%
%
%
\section{Type corrections and the value of $H_0$}
The zero-points needed for the type corrections are calculated
using Eq.~\ref{E21}. It was pointed out in Sect.~2.2 that $\log H_0$
should run horizontally in order to find an unbiased estimate for
$H_0$. In this section we look for such an slope.
Because the type-corrections depend on the adopted slope, this
fine-tuning of the slope must be carried out in an iterative manner.
This process consists of finding the type corrections 
$\Delta b'(\mathrm{T})$ for each test slope $a'$. Corrections are
made for both the field and calibrator samples. The process is
repeated until a horizontal $\langle\log H_0\rangle$ run is
found.

\begin{figure}
\resizebox{\hsize}{!}{\includegraphics[211,195][460,384]{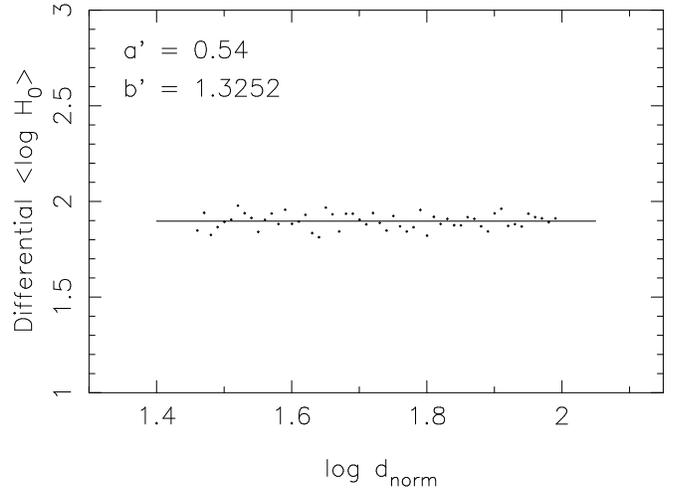}}
\caption{The differential $\langle\log H_0\rangle$ 
as a function of the log of normalized 
distance $\log d_\mathrm{norm}$
for the plateau galaxies with the adopted relation 
$\log V_\mathrm{max}=0.54\log D+1.325$. The solid line
is the average $\log H_0=1.897$.}
\label{F10}
\end{figure}

Consider first the diameter sample. When the criteria for the 
unbiased inverse plateau were imposed on the
sample, 2142 galaxies were left. For this subsample
the iteration yielded $a'=0.54$ (the straight line in 
Fig.~\ref{F10} is the least squares fit with a slope 0.003)
and when the corresponding
type corrections given in Table~1 were applied to the calibrator
sample and the slope forced to it one found 
$b'_\mathrm{cal}(6)=1.325$. 
The result is shown in Fig.~\ref{F10}.
The given inverse parameters predict an average
$\langle\log H_0\rangle=1.897$
(or $H_0=78.9\mathrm{\ km\,s^{-1}\,Mpc^{-1}}$). 

\begin{figure}
\resizebox{\hsize}{!}{\includegraphics[211,195][460,384]{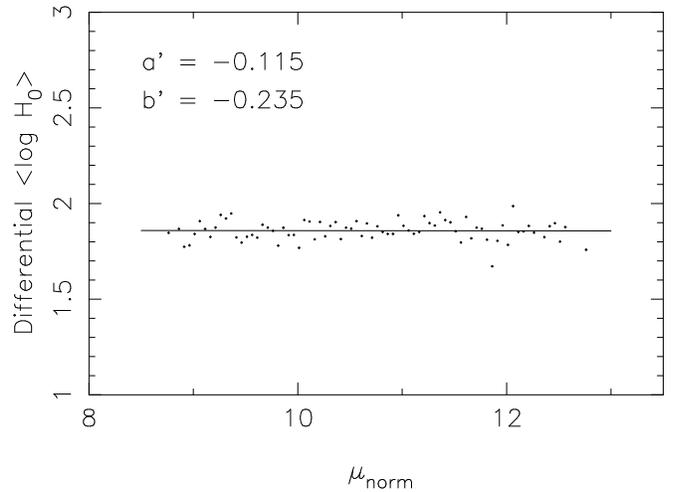}}
\caption{The differential $\langle\log H_0\rangle$ 
as a function of the log of normalized 
distance modulus $\mu_\mathrm{norm}$
for the plateau galaxies with the adopted relation 
$\log V_\mathrm{max}=-0.115M-0.235$. The solid line is
the average $\log H_0=1.869$.}
\label{F11}
\end{figure}

We treated the magnitude sample of 1713 galaxies in a similar
fashion. The resulting best fit is shown in Fig.~\ref{F11}.
The relevant slope is $a'=-0.115$ (the least squares fit yields
a slope 0.0004). The corresponding type corrections are given
in Table~1. The forced calibration gives $b'_\mathrm{cal}(6)=-0.235$. 
From this sample we find an average
$\langle\log H_0\rangle=1.869$ 
(or $H_0=72.4\mathrm{\ km\,s^{-1}\,Mpc^{-1}}$). 
In both cases the inverse estimates
for the Hubble constant ($H_0\approx80$ for the diameter relation
and $H_0\approx70$ for the magnitude relation) are considerably
larger than the corresponding estimates using
the direct Tully-Fisher relation ($H_0\approx55$).
%
\begin{table}
\label{T1}
\begin{center}
\begin{tabular}{ccc}
\hline
$\Delta b'(T)$&$a'=0.54$&$a'=-0.115$\\
\hline
$\Delta b'(1)$&0.125&0.110\\ 
$\Delta b'(2)$&0.156&0.124\\
$\Delta b'(3)$&0.129&0.096\\
$\Delta b'(4)$&0.095&0.058\\
$\Delta b'(5)$&0.069&0.030\\
$\Delta b'(6)$&0.0&0.0\\
$\Delta b'(7)$&-0.054&-0.042\\ 
$\Delta b'(8)$&-0.118&-0.075\\
\hline
\end{tabular}
\end{center}
\caption{The type corrections required for the relevant 
slopes $a'=0.54$ for the unbiased diameter sample and $a'=-0.115$
for the magnitude sample.}
\end{table}
%
%
%
%
%
%
%
%
%
%
%
\section{$H_0$ corrected for a calibrator selection bias}
The values of $H_0$ from the direct and inverse relations 
still disagree {\it even} after we have taken into account
the selection in $\log V_\mathrm{max}$, made the type corrections 
and used the relevant slope. 
There is -- however -- a serious possibility
left to explain the discrepancy. 
{\it The calibrator
sample used may not meet the theoretical requirements
of the inverse relation}. 
In order to transform the relative distance scale 
into an absolute one a properly chosen sample 
of calibrating galaxies is needed. 
What does ``properly chosen" mean? 
Consider first the direct relation for which 
it is essential to possess a 
calibrator sample, which is volume-limited for {\it each}
$p_\mathrm{cal}$. This means that for a $p_\mathrm{cal}$ 
one has $X_\mathrm{cal}$
which is drawn from the complete part of the gaussian distribution 
function $G(X;X_p,\sigma_{X_p})$, where the average $X_p=ap+b$.
If $\sigma_{X_p}$ is constant for all $p$ and the 
direct slope $a$ has been {\it correctly} derived 
from the unbiased field
sample, it will, when forced onto the calibrator sample,
bring about the correct calibrating zero-point.

As regards the calibration of the inverse relation the
sample mentioned above does not necessarily guarantee
a successful calibration. As pointed out by Teerikorpi
et al. (\cite{Teerikorpi99}) though the calibrator
sample is complete in the {\it direct} sense nothing has
been said about how the $p_\mathrm{cal}$'s relate to the 
corresponding cosmic distribution of $p$'s from which the
field sample was drawn. $\langle p\rangle_\mathrm{cal}$
should reflect the cosmic average $p_0$. 
If not, the relevant field slope when forced
to the calibrator sample will bring about a biased estimate
for $H_0$. 
Teerikorpi (\cite{Teerikorpi90}) 
already recognized that this
could be a serious problem. He studied, however,
clusters of galaxies where a nearby (calibrator) cluster
obeys a different slope than a distant cluster.
Teerikorpi et al. (\cite{Teerikorpi99}) 
developed the ideas further
and showed how this problem may be met also when using
field galaxies. The mentioned bias when using the relevant
slope can be corrected
for but is a rather complicated task. For the theoretical
background of the ``calibrator selection bias" 
consult Teerikorpi et al. (\cite{Teerikorpi99}).

One may -- as pointed out by Teerikorpi et al.
(\cite{Teerikorpi99}) -- use instead of the relevant
slope the calibrator slope which also predicts a biased
estimate for $H_0$ but which can be corrected for in
a rather straightforward manner.
For the diameter relation the average correction 
term reads as
%
\begin{equation}
\label{E23}
\\ \Delta\log H_0=(3-\alpha)\,\ln 10\,\sigma_D^2
\times\left[\frac{a'_\mathrm{cal}}{a'}-1\right],
\end{equation}
where $\sigma_D$ is the dispersion of the log linear
diameter $\log D_\mathrm{linear}$ and $\alpha$ gives
the radial number density gradient : $\alpha=0$ corresponds to
a strictly homogeneous distribution of galaxies.
For magnitudes the correction term follows from
(cf. Teerikorpi \cite{Teerikorpi90})
%
\begin{equation}
\label{E24}
\\ \Delta\log H_0=
0.2\,\left[\frac{a'_\mathrm{cal}}{a'}-1\right]
\times(\langle M\rangle-M_0).
\end{equation}
Because $\langle M\rangle-M_0$ simply reflects the classical
Malmquist bias one finds:
%
\begin{equation}
\label{E25}
\\ \Delta\log H_0=\frac{(3-\alpha)\ln 10}{5}\,\sigma_M^2
\times0.2\left[\frac{a'_\mathrm{cal}}{a'}-1\right],
\end{equation}
Note that one may use the calibrator slope and consequently
the correction formulas {\it irrespective} of the nature
of the calibrator sample 
(Teerikorpi et al. \cite{Teerikorpi99}). If the calibrator
sample would meet the requirement mentioned, the value corrected with
Eqs.~\ref{E23} or~\ref{E25} should equal values obtained
from the relevant slopes. Furthermore, our analysis carried
out so far would have yielded an unbiased estimate for $H_0$
and thus the problems would be in the direct analysis.
However, if the requirement is not met one should prefer the
corrective method using the calibrator slope.
%
%
%
\begin{figure}
\resizebox{\hsize}{!}{\includegraphics[211,195][460,384]{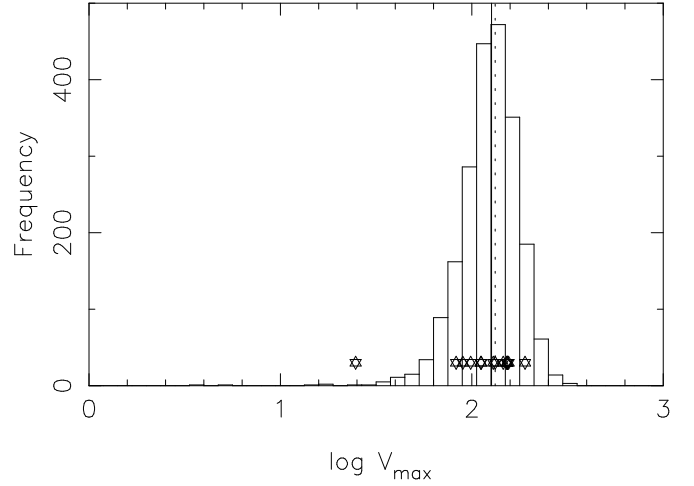}}
\caption{Histogram of the 
$\log V_\mathrm{max}$ values and the 
individual calibrators (labelled with stars). 
The vertical solid line gives the median of 
the plateau 
$\mathrm{Med}(\log V_\mathrm{max}^\mathrm{plateau})=2.10$
and the dotted line gives the median of the calibrators 
$\mathrm{Med}(\log V_\mathrm{max}^\mathrm{calib})=2.11$.}
\label{F12}
\end{figure}
%
%
%
\begin{figure}
\resizebox{\hsize}{!}{\includegraphics[211,195][460,384]{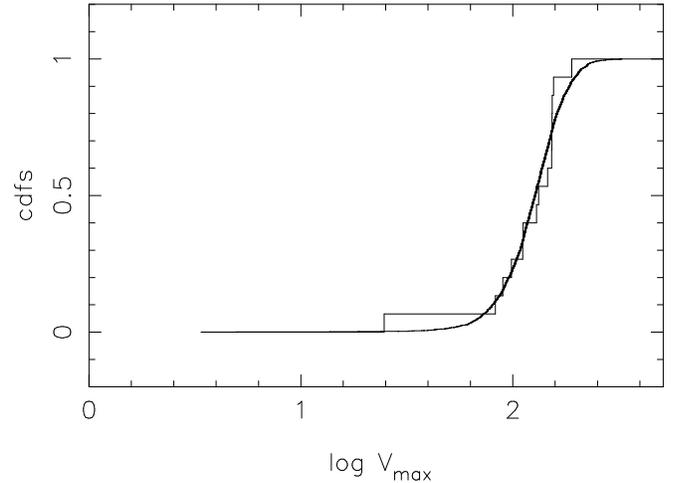}}
\caption{The Kolmogorov-Smirnov test for the diameter sample.
Pay attention to the rather remarkable similarity between 
the cumulative distribution functions (cdfs).}
\label{F13}
\end{figure}
%
%
%
\begin{figure}
\resizebox{\hsize}{!}{\includegraphics[211,195][460,384]{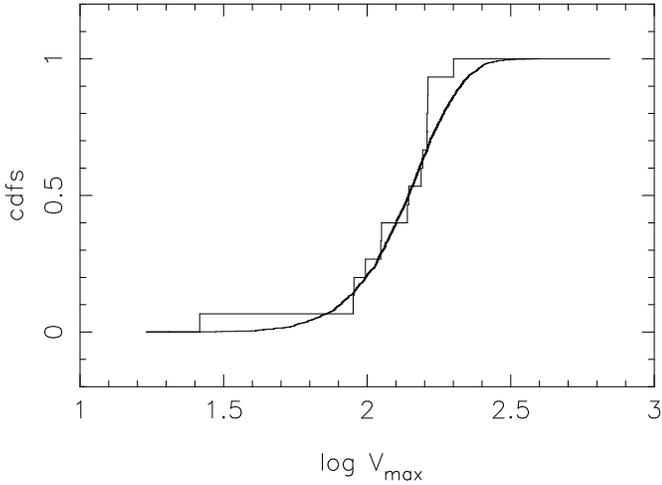}}
\caption{The Kolmogorov-Smirnov test for the magnitude sample. Again
the cdfs are quite similar.}
\label{F14}
\end{figure}
\subsection{Is the calibrator sample representative?}
Is the calibrator bias present in our case? Recall that the
calibrators used were sampled from the nearby field
to have high quality distance moduli mostly from the 
HST Cepheid measurements. 
This means that we have no a priori guarantee that the
calibrator sample used will meet the criterium
required.
We compare the type-corrected diameter and magnitude
samples with the calibrator sample. Note that for the
diameter sample we use only galaxies residing in the
unbiased inverse plateau (i.e. the small selection effect
in $\log V_\mathrm{max}$ has been eliminated).
In Fig.~\ref{F12} we show the histogram of the
$\log V_\mathrm{max}$ values for the diameter sample
and the individual calibrators (labelled as stars). 
The vertical solid line gives the median of 
the plateau 
$\mathrm{Med}(\log V_\mathrm{max}^\mathrm{plateau})=2.10$
and the dotted line gives the median of the calibrators 
$\mathrm{Med}(\log V_\mathrm{max}^\mathrm{calib})=2.11$.
In the case of magnitudes both the field and calibrator
sample have the same median (2.14).
The average values for the diameter case were
$\langle\log V_\mathrm{max}^\mathrm{plateau}\rangle=2.09$
and
$\langle\log V_\mathrm{max}^\mathrm{calib}\rangle=2.06$,
and for the magnitude case
$\langle\log V_\mathrm{max}^\mathrm{mag}\rangle=2.12$
and
$\langle\log V_\mathrm{max}^\mathrm{calib}\rangle=2.08$. 
Both the diameter and the magnitude field samples were 
subjected to strict limits, which means that both 
inevitably suffer from the classical Malmquist bias.
In order to have a representative calibrator sample in the sense
described, we would have expected a clear difference
between the field and calibrator samples. That the statistics 
are very close to each other lends credence to the assumption 
that the calibrator selection bias is present. 

We also made tests using the Kolmogorov-Smirnov 
statistics (Figs.~\ref{F13} and~\ref{F14}). 
In this test a low significance level
should be considered as counterevidence for a hypothesis that two
samples rise from the same underlying distribution. 
We found relatively high significance
levels (0.89 for the diameter sample and 0.3 for the magnitude sample).
Neither these findings corroborate the 
hypothesis that the calibrator sample is drawn from the cosmic
distribution and hence the use of Eqs.~\ref{E23} or~\ref{E25}
is warranted.
\subsection{The dispersion in $\log D_\mathrm{linear}$}
%
%
%
%
\begin{figure}
\resizebox{\hsize}{!}{\includegraphics[211,195][460,384]{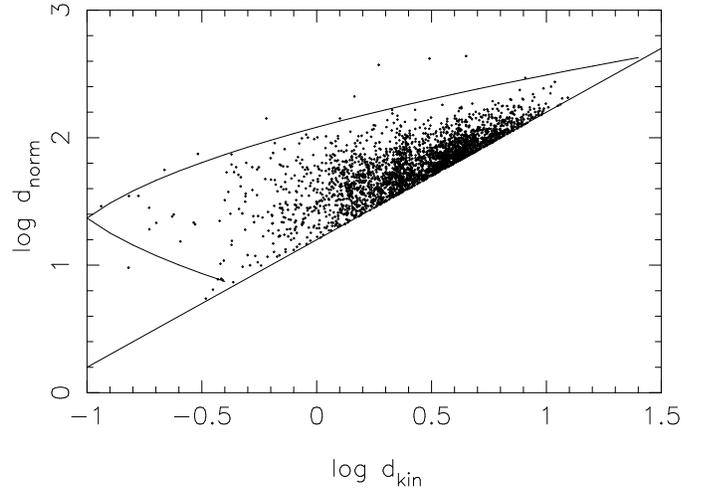}}
\caption{A classical Spaenhauer diagram for normalized distances vs.
kinematical distances with a presumed dispersion $\sigma_X=0.28$.
}
\label{F15}
\end{figure}
%
%
%
%
\begin{figure}
\resizebox{\hsize}{!}{\includegraphics[211,195][460,384]{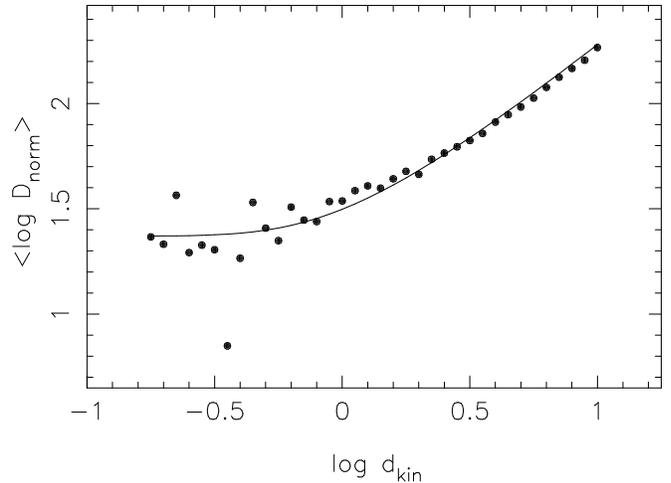}}
\caption{Comparison between average values of 
$\langle\log d_\mathrm{norm}\rangle$
for different kinematical distances and the theoretical prediction
calculated from Eq.~\ref{E26} with $X_0^*=1.37$ and $\sigma_X=0.28$.
}
\label{F16}
\end{figure}
In order to find a working value for the dispersion in 
$\log D_\mathrm{linear}$, we first consider the classical
Spaenhauer diagram (cf. Sandage \cite{Sandage94a}, 
\cite{Sandage94b}).
In the Spaenhauer diagram one studies the behaviour of 
$X$ as a function of the redshift. If the observed 
redshift could be translated into the corresponding cosmological distance, 
then $X$ inferred from $x$ and the redshift would genuinely 
reflect the true size of a galaxy.

In practice, the observed redshift cannot be considered as
a direct indicator of the cosmological distance because of the
inhomogeneity of the Local Universe. Peculiar motions
should also be considered. Thus the inferred $X$
suffers from uncertainties in the underlying kinematical model.
The Spaenhauer diagram as a diagnostics for the distribution function
is always constrained by our knowledge of the form of the true
velocity-distance law.

Because the normalized distance (cf. Sect.~3.) is proportional
to the linear diameter we construct the Spaenhauer diagram as
$\log d_\mathrm{norm}$ vs. $\log d_\mathrm{kin}$ thus avoiding the
uncertainties in the absolute distance scale. The problems with
relative distance scale are -- of course -- still present.
The fit shown in Fig.~\ref{F15} is not unacceptable.
The dispersion used was $\sigma_X=0.28$, a value inferred 
from the dispersion in absolute B-band magnitudes $\sigma_M=1.4$
(Fouqu\'e et al. \cite{Fouque90}) 
based on the expectation that the
dispersion in log linear diameter should be one fifth of that
of absolute magnitudes. 
 
We also looked how the average values 
$\langle\log d_\mathrm{norm}\rangle$
at different kinematical distances compare to the theoretical
prediction which, in a strictly limited sample of $X$'s, 
at each log distance is formally expressed as
%
\begin{equation}
\label{E26}
\\  \langle X\rangle_\mathrm{d} = X_0^*+ 
\frac {2\sigma_X}{\sqrt2\pi}
\frac
{\exp\bigl[-(X_\mathrm{lim}-X_0^*)^2/(2\sigma_X^2)\bigr]}
{\mathrm{erfc}\bigl[(X_\mathrm{lim}-X_0^*)/(\sqrt2\sigma_X)\bigr]}.
\end{equation}
Here $X$ refers to $\log d_\mathrm{norm}$.
The curve in Fig.~\ref{F16} is based on $X_0^*=1.37$
and $\sigma_X=0.28$. The averages from the data are shown as bullets. 
The data points follow the theoretical prediction reasonably well.
\subsection{Corrections and the value of $H_0$}
%
%
%
\begin{figure}
\resizebox{\hsize}{!}{\includegraphics[211,195][460,384]{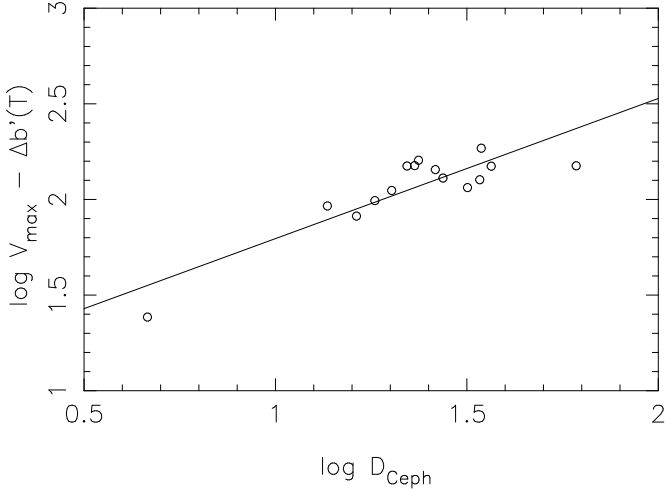}}
\caption{A least squares fit the type corrected calibrator
sample yielding $a'=0.73$.
The type correction was based on $a'=0.54$.}
\label{F17}
\end{figure}
%
%
%
\begin{figure}
\resizebox{\hsize}{!}{\includegraphics[211,195][460,384]{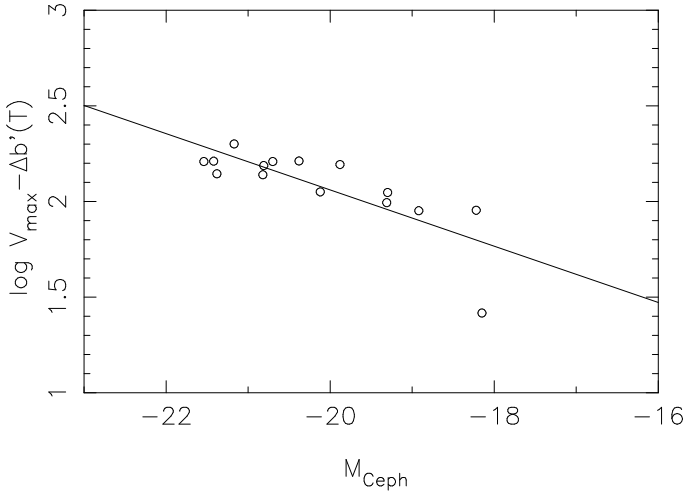}}
\caption{A least squares fit the type corrected calibrator
sample yielding $a'=-0.147$.
The type correction was based on $a'=-0.115$.}
\label{F18}
\end{figure}
Consider a strictly homogeneous universe, i.e. $\alpha=0$.
In Eqs. \ref{E23} and \ref{E25} one needs
values for slope $a'_\mathrm{c}$. 
Least squares fit to the type-corrected calibrator sample yields
$a'_\mathrm{c}=0.73$ for the diameter relation
and $a'_\mathrm{c}=-0.147$ for the magnitude relation.
(cf. Figs. \ref{F17} and \ref{F18}).
These slopes correspond to diameter zero-point
$b'_\mathrm{c}(6)=1.066\pm0.103$ 
and to magnitude zero-point
$b'_\mathrm{c}=-0.879\pm0.131$ 
The {\it biased} estimates for average $\log H_0$ are
$\langle\log H_0\rangle = 1.910\pm0.188$ for the diameters
and $\langle\log H_0\rangle = 1.876\pm0.176$ for the magnitudes.
For the zero-points and the averages we have given the $1\sigma$
standard deviations. The {\it mean error} in the averages is
estimated from
%
\begin{equation}
\label{E27}
\\ \epsilon_{\langle\log H_0\rangle}\approx
\sqrt{\frac{\sigma_{B'}^2}{N_\mathrm{cal}}
+\frac{\sigma_{\log H_0}^2}{N_\mathrm{gal}}},
\end{equation}
where $\sigma_{B'}=\sigma_{b'}/a'_\mathrm{cal}$ for
diameters and $\sigma_{B'}=0.2\sigma_{b'}/a'_\mathrm{cal}$
for magnitudes. The use of Eq.~\ref{E27} is acceptable
because the dispersion in $b'$ does not correlate with
the dispersion $\log H_0$. With the given slopes and dispersions
we find:
\begin{itemize}
\item $\langle\log H_0\rangle = 1.910\pm0.037$ for the diameters
\item $\langle\log H_0\rangle = 1.876\pm0.046$ for the magnitudes.
\end{itemize}

Eq. \ref{E23} predicts an average correction term for
the slopes $a'_\mathrm{c}=0.73$ and $a'=0.54$ 
together with $\sigma_X=0.28$ $\Delta\log H_0=0.191$. 
and Eq. \ref{E25} with $a'_\mathrm{c}=-0.147$,$a'=-0.115$ 
and $\sigma_M=1.4$ $\Delta\log H_0=0.151$. When applied
to the above values we get the corrected, unbiased estimates
\begin{itemize}
\item $\langle\log H_0\rangle= 1.719\pm0.037$ for the diameters
\item $\langle\log H_0\rangle = 1.725\pm0.046$ for the magnitudes.
\end{itemize}
These values translate into Hubble constants
\begin{itemize}
\item $H_0=52^{+5}_{-4}\mathrm{\ km\,s^{-1}\,Mpc^{-1}}$ 
for the inverse diameter
B-band Tully-Fisher relation, and
\item $H_0=53^{+6}_{-5}\mathrm{\ km\,s^{-1}\,Mpc^{-1}}$ 
for the inverse magnitude
B-band Tully-Fisher relation.
\end{itemize}
These corrected values are in good concordance with each
other as well as with the estimates established from
the direct diameter Tully-Fisher relation 
(Theureau et al. \cite{Theureau97b}).
Note that the errors in the magnitude relation are slightly
larger than in the diameter relation. This is expected because
for the diameter relation we possess more galaxies. The error
is however mainly governed by the uncertainty in the calibrated
zero-point. This is expected because though the dispersion in
inverse relation as such is large it is compensated by the
number galaxies available.

Finally, how significant an error do the correction formulae
induce? We suspect the error to mainly depend on $\alpha$.
The correction above was based on the assumption of
homogeneity (i.e. $\alpha=0$). Recently 
Teerikorpi et al. (\cite{Teerikorpi98}) found
evidence that the average density radially 
decreases around us ($\alpha\approx0.8$)
confirming the more general (fractal) analysis 
by Di Nella et al. (\cite{DiNella96}).
Using this value of $\alpha$
we find $\Delta\log H_0=0.140$ for the diameters
and $\Delta\log H_0=0.111$ for the magnitudes
yielding
\begin{itemize}
\item $\langle\log H_0\rangle = 1.770\pm0.037$ for the diameters
\item $\langle\log H_0\rangle = 1.765\pm0.046$ for the magnitudes.
\end{itemize}
In terms of the Hubble constant we find
\begin{itemize}
\item $H_0=59^{+5}_{-4}\mathrm{\ km\,s^{-1}\,Mpc^{-1}}$ 
for the inverse diameter
B-band Tully-Fisher relation, and
\item $H_0=58^{+6}_{-5}\mathrm{\ km\,s^{-1}\,Mpc^{-1}}$ 
for the inverse magnitude
B-band Tully-Fisher relation.
\end{itemize}
%
%
%
%
%
\section{Summary}
In the present paper we have examined how to apply 
the inverse Tully-Fisher relation to the problem of 
determining the value of the Hubble constant,
$H_0$, in the practical context of the large galaxy sample KLUN.
We found out that the implementation
of the inverse relation is not as simple task as one
might expect from the general considerations (in particular the quite
famous result of the unbiased nature of the relation).
We summarize our main results as follows.

\begin{enumerate}
\item A straightforward application of the inverse relation
consists of finding the average Hubble ratio for each kinematical
distance and tranforming the relative distance into an absolute one
through calibration. The 15 calibrator galaxies used 
were drawn from the field
with cepheid distance moduli obtained mostly from the HST
observations. The inverse diameter relation predicted
$H_0\approx 80\mathrm{\ km\,s^{-1}\,Mpc^{-1}}$  
and the magnitude relation predicted
$H_0\approx 70\mathrm{\ km\,s^{-1}\,Mpc^{-1}}$  
The diameter value for $H_0$ is about 50 percent and
the magnitude value about 30 percent larger than those
obtained from the direct relation
(cf. Theureau et al. \cite{Theureau97b}).

\item We examined whether this discrepancy could be resolved 
in terms of some selection effect in $\log V_\mathrm{max}$
and the type dependence of the zero-points on the Hubble type.
One expects these to have some influence on the
derived value of $H_0$. Only a minuscule effect was observed.

\item There is -- however -- a new kind of bias involved:
if the $\log V_\mathrm{max}$-distribution
of the calibrators does not reflect the cosmic distribution
of the field sample {\it and} the relevant slope for the field galaxies
differs from the calibrator slope the average value of $\log H_0$ 
will be biased if the relevant slope is used
(Teerikorpi et al. \cite{Teerikorpi99}).

\item We showed for the unbiased inverse plateau galaxies
i.e. a sample without galaxies probably suffering from selection in
$\log V_\mathrm{max}$,
that the calibrators and the field sample obey different inverse 
diameter slopes, namely $a'_\mathrm{cal}=0.73$ and $a'= 0.54$,
Also, the magnitude slopes differed from each other
($a'_\mathrm{cal}=-0.147$ and $a'= -0.115$). For the
diameter relation we were able to use 2142 galaxies and for
the magnitude relation 1713 galaxies. These sizes are
significant.

\item We also found evidence that the calibrator sample
does {\it not} follow the cosmic distribution of 
$\log V_\mathrm{max}$ for the field galaxies. This means
that if the relevant slopes are used a too large value for $H_0$
is found. Formally, this calibrator selection bias could be
corrected for but is a complicated task. 

\item One may use instead of the relevant slope the calibrator
slope which also brings about a biased value of $H_0$. Now,
however, the correction for the bias is an easy task.
Furthermore, this approach can be used irrespective of the nature
of the calibrator sample and should yield an unbiased estimate
for $H_0$.

\item When we adopted this line of approach we found
\begin{itemize}
\item $H_0=52^{+5}_{-4}\mathrm{\ km\,s^{-1}\,Mpc^{-1}}$ 
for the inverse diameter
B-band Tully-Fisher relation, and
\item $H_0=53^{+6}_{-5}\mathrm{\ km\,s^{-1}\,Mpc^{-1}}$ 
for the inverse magnitude
B-band Tully-Fisher relation
\end{itemize}
for a strictly homogeneous distribution of galaxies ($\alpha=0$) and
\begin{itemize}
\item $H_0=59^{+5}_{-4}\mathrm{\ km\,s^{-1}\,Mpc^{-1}}$ 
for the inverse diameter
B-band Tully-Fisher relation, and
\item $H_0=58^{+6}_{-5}\mathrm{\ km\,s^{-1}\,Mpc^{-1}}$ 
for the inverse magnitude
B-band Tully-Fisher relation.
\end{itemize}
for a decreasing radial density 
gradient ($\alpha=0.8$).
\end{enumerate}

These values are in good concordance with each other as well as
with the values established from the corresponding direct 
Tully-Fisher relations derived  by Theureau et al. (\cite{Theureau97b}),
who gave a strong case for
the long cosmological distance scale consistently supported by
Sandage and his collaborators. Our analysis also 
establishes a case supporting such a scale.
It is worth noting that this is the first time when the
{\it inverse} Tully-Fisher relation clearly lends credence
to small values of the Hubble constant $H_0$.
\begin{acknowledgements}
{We have made use of data from the Lyon-Meudon extragalactic
Database (LEDA) compiled by the LEDA team at the CRAL-Observatoire
de Lyon (France). This work has been supported by the Academy of Finland
(projects ``Cosmology in the Local Galaxy Universe'' and
``Galaxy streams and structures in the Local Universe'').
T. E. would like to thank G. Paturel and his staff for hospitality
during his stay at the Observatory of Lyon in May 1998. Finally,
we thank the referee for useful comments and constructive
criticism.}
\end{acknowledgements}
\appendix
\section{The relevant slope and an unbiased $H_0$}
In this appendix we in simple manner demonstrate how the
relevant slope introduced in Sect. 2.2 indeed is the
slope to be used. Consider
%
\begin{equation}
\label{EA1}
\\ \log H_0 = \log V_\mathrm{cor}-\log R_\mathrm{iTF},
\end{equation}
where the velocity corrected for the peculiar motions, $V_\mathrm{cor}$,
depends on the relative kinematical distance scale as 
%
\begin{equation}
\label{EA2}
\\ \log V_\mathrm{cor}=\log C_1+\log d_\mathrm{kin}
\end{equation}
and the inverse Tully-Fisher distance in Mpc is\footnote{
The numerical constant $\beta=1.536274$
connects $x$ in 0.1 arcsecs, $X$ in kpc and
$R_\mathrm{iTF}$ in Mpc} 
%
\begin{equation}
\label{EA3}
\\ \log R_\mathrm{iTF}= Ap+B_\mathrm{cal}-x+\beta
\end{equation}
The constant $C_1$ was defined by
Eq.~\ref{E11} and can be decomposed into
$\log C_1 = \log H_0^*+\log C_2$. $H_0^*$ is the true value of the Hubble
constant and $C_2$ transforms the relative distance scale into the
absolute one: $\log R_\mathrm{kin}=\log d_\mathrm{kin}+\log C_2$.
Because $X_\mathrm{kin}=\log R_\mathrm{kin}+x-\beta$
Eq.~\ref{EA1} reads:
%
\begin{equation}
\label{EA4}
\\ \log H_0-\log H_0^*
=X_\mathrm{kin}-Ap-B_\mathrm{cal}.
\end{equation}
Consider now a subsample of galaxies at a constant $X_\mathrm{o}$.
By realizing that $X_\mathrm{kin}=X_\mathrm{o}+(B'-B_\mathrm{in})$, where
$B'$ gives the true distance scale and $B_\mathrm{in}$ depends on the adopted
distance scale (based on the input $H_0$), and by 
taking the average over $X_\mathrm{o}$ Eq.~\ref{EA4} yields
%
\begin{equation}
\label{EA5}
\\ \langle\log H_0\rangle_{X_\mathrm{o}}-\log H_0^*
=X_\mathrm{kin}-A\langle p\rangle_{X_\mathrm{o}}-B_\mathrm{cal}.
\end{equation}
The use of $B'$ is based on two presumptions, 
namely that the underlying kinematical
model indeed brings about the correct relative distance scale and that the
adopted value for $C_1$ genuinely reflects the true absolute distance scale.
If the adopted slope $a'_\mathrm{o}$ is the relevant one we find using 
Eq.~\ref{E19}
$A\langle p\rangle_{X_\mathrm{o}}=X_\mathrm{o}-B_\mathrm{in}$ and
%
\begin{equation}
\label{EA6}
\\ \langle\log H_0\rangle_{X_\mathrm{o}}-\log H_0^*
=(B'-B_\mathrm{in})-(B_\mathrm{cal}-B_\mathrm{in}).
\end{equation}
As a final result we find
%
\begin{equation}
\label{EA7}
\\ \langle\log H_0\rangle_{X_\mathrm{o}}-\log H_0^*
= (b'_\mathrm{cal}-b'_\mathrm{true})/a'_\mathrm{o}.
\end{equation}
Because Eq.~\ref{EA7} is valid for each $X_\mathrm{o}$, 
the use of the relevant slope
necessarily guarantees a horizontal run for $\langle\log H_0\rangle$ 
as a function of $X_\mathrm{o}$.
\section{Note on a theoretical diameter slope $a'\sim0.75$}
Theureau et al. (\cite{Theureau97a}) presented theoretical arguments
which supported the inverse slope $a'=0.5$ being derived from the field
galaxies. Consider a pure rotating disk
(the Hubble type 8). The square of the rotational velocity measured at
the radius $r_\mathrm{max}$ at which the rotation has its maximum is
directly proportional to the mass within $r_\mathrm{max}$, which in turn
is proportional to the square of $r_\mathrm{max}$.
Hence, $\log V_\mathrm{max}\propto0.5\log r_\mathrm{max}$. By adding
a bulge with a mass-to-luminosity ratio differing from that of the disk,
and a dark halo with mass proportional to the luminous mass, one can as
a first approximation understand the dependence of the zero-point of
the inverse relation on the Hubble type.

However, the present study seems to require that the theoretical
slope $a'$ is closer to 0.75 rather than 0.5. 
The question arises whether the
simple model used by Theureau et al. (\cite{Theureau97a}) 
could in some natural
way be revised in order to produce a steeper slope. In fact, the model
assumed that for each Hubble type the mass-to-luminosity ratio
$M/L$ is constant in galaxies of different sizes (luminosities). If one
allows $M/L$ to depend on luminosity, the slope $a'$ will differ from
0.5. Especially, if $M/L\propto L^{0.25}$, one may show that the model
predicts the inverse slope $a'=0.75$. The required luminosity dependence
of $M/L$ is interestingly similar to that of the fundamental plane for
elliptical galaxies and bulges 
(Burstein et al. \cite{Burstein97}). 
The questions of the slope, the
mass-to-luminosity ratio and type-dependence will be investigated elsewhere
by Hanski \& Teerikorpi (1999, in preparation).
\section{How to explain $a'_\mathrm{obs}<a'$?}
%
%
%
\begin{figure}
\resizebox{\hsize}{!}{\includegraphics[211,195][460,384]{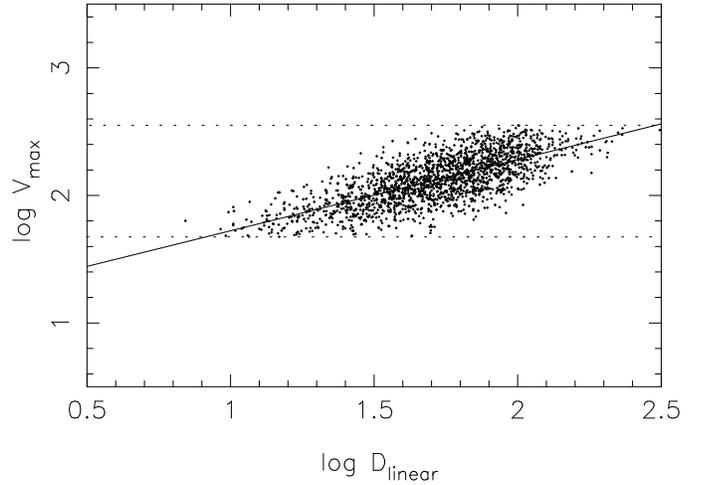}}
\caption{
A synthetic Virgo supercluster subjected to a bias caused by measurement
errors in apparent diameters and
upper and lower cut-offs in $\log V_\mathrm{max}$.
}
\label{FC1}
\end{figure}
Among other things ET97 discussed how a gaussian measurement error
$\sigma_x$ in apparent diameters
yields a too large value for $H_0$.
How does the combination of cut-offs in $\log V_\mathrm{max}$ -distribution
and this bias affect the slope? We examined this problem by
using a synthetic Virgo Supercluster 
(cf. Ekholm \cite{Ekholm96}). As a luminosity
function we chose 
%
\begin{equation}
\label{EC1}
\\ \log D=0.28\times G(0,1)+1.2
\end{equation}
and as the inverse relation
%
\begin{equation}
\label{EC2}
\\ \log V_\mathrm{max}=0.11\times G(0,1)+a'_\mathrm{t}\log D
+0.9, 
\end{equation}
where $G(0,1)$ refers to a normalized gaussian random variable. 
As the ``true'' inverse slope we used $a'_\mathrm{t}=0.75$. The
other numerical values were adjusted in order 
to have a superficial resemblance with Fig.~\ref{F5}. 
We first subjected the synthetic
sample to the upper and lower cut-offs 
in $\log V_\mathrm{max}$ given in Sect.~4.
The resulting slope was $a'=0.692$. A dispersion of
$\sigma_x=0.05$ yielded $a'=0.642$ and $\sigma_x=0.1$
$a'=0.559$. The inverse Tully-Fisher diagram for the latter case is
shown in Fig.~\ref{FC1}. 
Though the model for the errors is rather
simplistic this experiment shows a natural way of flattening the
observed slope $a'$ with respect to the input slope $a'_\mathrm{t}$.
\end{document}